\documentclass[fleqn,usenatbib]{mnras}

\usepackage[T1]{fontenc}

\DeclareRobustCommand{\VAN}[3]{#2}
\let\VANthebibliography\thebibliography
\def\thebibliography{\DeclareRobustCommand{\VAN}[3]{##3}\VANthebibliography}

\usepackage{graphicx}	% Including figure files
\usepackage{amsmath}	% Advanced maths commands
\usepackage{amssymb}	% Extra maths symbols
\usepackage{esdiff}
\usepackage{physics}
\usepackage{newtxtext,newtxmath}

\title[Puzzling large-scale polarization in A523]{Puzzling large-scale polarization in the galaxy cluster Abell 523}

\author[V. Vacca et al.]{
Valentina Vacca$^{1}$\thanks{E-mail: valentina.vacca@inaf.it},
Federica Govoni$^{1}$,
Matteo Murgia$^{1}$,
Rick Perley$^{2}$,
Luigina Feretti$^{3}$,
\newauthor 
Gabriele Giovannini$^{3,4}$,
Ettore Carretti$^{3}$,
Fabio Gastaldello$^{5}$,
Filippo Cova$^{5}$,
Paolo Marchegiani$^{1,6}$,
\newauthor
Elia Battistelli$^{7}$,
Walter Boschin$^{8,9,10}$,
Torsten En{\ss}lin$^{11}$,
Marisa Girardi$^{12}$,
Francesca Loi$^{1}$,
Federico Radiconi$^{7}$\\
$^{1}$ \quad INAF-Osservatorio
Astronomico di Cagliari, Via della Scienza 5, I-09047 Selargius (CA), Italy\\
$^{2}$ \quad National Radio Astronomy Observatory, P.O.Box O, Socorro, NM, 87801\\
$^{3}$ \quad INAF - Istituto di Radioastronomia, Via P. Gobetti 101, 40129 Bologna, Italy\\
$^{4}$ \quad Dipartimento di Fisica e Astronomia, Universita di Bologna, Via Gobetti 93/2, 40122, Bologna, Italy\\
$^{5}$ \quad IASF - Milano, INAF, Via Corti 12, I-20133 Milan, Italy\\
$^{6}$ \quad Wits Centre for Astrophysics, School of Physics, University of the Witwatersrand, 1 Jan Smuts Avenue, Johannesburg 2050, South Africa\\
$^{7}$ \quad Sapienza—University of Rome—Physics department, Piazzale Aldo Moro 5—I-00185, Rome, Italy\\
$^{8}$ \quad Fundaci\'on G. Galilei - INAF (Telescopio Nazionale Galileo), Rambla J. A. Fern\'andez P\'erez 7, E-38712 Bre$\tilde{n}$a Baja (La Palma), Spain\\
$^{9}$ \quad Instituto de Astrof\'isica de Canarias, C/V\'ia Lactea s/n, E-38205 La Laguna (Tenerife), Spain\\
$^{10}$ \quad Departamento de Astrof\'isica, Univ. de La Laguna, Av. del Astrof\'isico Francisco S\'anchez, s/n, E-38205 La Laguna (Tenerife), Spain\\
$^{11}$ \quad Max Planck Institute for Astrophysics, Karl-Schwarzschildstr. 1, 85741 Garching, Germany\\
$^{12}$ \quad Dipartimento di Fisica dell’Università degli Studi di Trieste - Sezione di Astronomia, via Tiepolo 11, I-34143 Trieste, Italy\\
}

\date{Accepted XXX. Received YYY; in original form ZZZ}

\pubyear{2015}

\begin{document}
\label{firstpage}
\pagerange{\pageref{firstpage}--\pageref{lastpage}}
\maketitle

\begin{abstract}
Large-scale magnetic fields reveal themselves through diffuse synchrotron sources observed in galaxy clusters such as radio halos. Total intensity filaments of these sources have been observed in polarization as well, but only in three radio halos out of about one hundred currently known. 
In this paper we analyze new polarimetric Very Large Array data of the diffuse emission in the galaxy cluster Abell 523 in the frequency range 1-2\,GHz. We find for the first time evidence of polarized emission 
on scales of $\sim$2.5\,Mpc. 
Total intensity emission is observed only in the central part of the source, likely due to observational limitations.  
 To look for total intensity emission beyond the central region, we combine these data with single-dish observations from the Sardinia Radio Telescope and we compare them with multi-frequency total intensity observations obtained with different instruments, including the LOw Frequency ARray and the Murchison Widefield Array. By analysing the rotation measure properties of
the system and utilizing numerical simulations, we infer that this polarized emission is associated with filaments of the radio halo located in the outskirts of the system, in the peripheral region closest to the observer.
\end{abstract}

%%%%%%%%%%%%%%%%%%%%%%%%%%%%%%%%%%%%%%%%%%%%%%%%%%

%%%%%%%%%%%%%%%%% BODY OF PAPER %%%%%%%%%%%%%%%%%%

\section{Introduction}
Radio halos are diffuse synchrotron sources  detected in about 100 galaxy clusters hosting merger phenomena \citep[see, e.g., ][]{Feretti2012,vanWeeren2019}.
They are characterized by a size of $\sim$1-2\,Mpc, low
radio brightness ($\sim$0.1\,$\mu$Jy\,arcsec$^{-2}$ at 1.4-GHz) and steep-spectra ($S_{\nu}\,\propto\,\nu^{-\alpha}$, with $\alpha\,\simeq\,1-1.4$). These sources are typically unpolarized, likely because of internal depolarization of the signal, sensitivity and spatial and spectral resolution limitations \citep{Govoni2013}. Total intensity and polarized emission from these sources represent a valuable probe sampling the central cluster volume that allows us to study intracluster medium (ICM) properties with an high level of accuracy 
\citep{Vacca2010}. Polarimetric observations of radio halos provide information on the strength and structure of the ICM magnetization relatively independently from the synchrotron radiation emitting relativistic electrons.

Background polarized radio emission passing through a magneto-ionic medium of spatial depth $L$  undergoes a rotation of the polarization angle $\Psi(\lambda)$ versus the squared observing wavelength $\lambda^2$, leading to an observed polarization angle of
\begin{equation}
    \Psi(\lambda)=\Psi_0+\mathrm{RM}\lambda^2,
    \label{eq1}
\end{equation}
where $\Psi_0$ is the intrinsic polarization angle of the initial radiation and RM the rotation measure defined as the integral 
of the thermal gas density ($n_{\rm e}$) times the magnetic field component along the line of sight ($B_{\parallel}$) 
\begin{equation}
    \frac{\mathrm{RM}}{{\rm rad\,m^{-2}}}=812\int_0^{ L/{\rm [kpc]}}\frac{n_{\rm e}(l^{\prime})}{\rm [cm^{-3}]}\frac{B_{\rm \parallel}(l^{\prime})}{\rm [\mu G]} \mathrm{d}l^{\prime}.
    \label{eq2}
\end{equation}
If the radio emitting plasma is mixed with the magneto-ionic medium, as seems to be the case in radio halos, 
each emitting volume along the line of sight suffers a different Faraday rotation and the corresponding polarization angle
varies according to the relation in equation\,\ref{eq1}. This translates into depolarization for the observer and a 
deviation from the $\lambda^2$ proportionality of the overall observed polarisation angle.
To describe this scenario \cite{Burn1966} introduced the Faraday depth $\phi(l)$
\begin{equation}
    \frac{\phi(l)}{\rm rad\,m^{-2}}=812\int_0^{ l/{\rm [kpc]}}\frac{n_{\rm e}(l^{\prime})}{\rm [cm^{-3}]}\frac{B_{\rm \parallel}(l^{\prime})}{\rm [\mu G]}\mathrm{d}l^{\prime}.
    \label{eq3}
\end{equation}
Here, $l^{\prime}$ is the distance along the line of sight at which the emitting volume is located with respect to the observer. 
Only in the case of
a completely external Faraday rotating screen will the Faraday depth and
rotation measure be the same.
For sources located in the peripheral region of a galaxy cluster, the internal depolarization is expected to be low due to the low thermal gas density and weak magnetic field there. Therefore, as shown by \cite{Loi2019a}, polarized filaments of diffuse emission are easier to detect at large distance from the cluster centre ($\sim$1.5\,Mpc). To date, filaments of polarized emission have been observed with high significance only in three galaxy clusters (Abell 2255, \citealt{Govoni2005,Pizzo2011}; MACS J0717.5+3745, \citealt{Bonafede2009}; and Abell 523, \citealt{Girardi2016}) and polarization at $<$5$\sigma$ level has been reported for the radio halo in 1E 0657-55.8 \citep{Shimwell2014}. Thanks to the improved performance, the new generation of radio interferometers should enable the detection of  polarized emission in radio halos in a larger number of galaxy clusters, at least for intermediate-high power radio halos $P_{\rm 1.4\,GHz}\gtrsim 10^{24}$\,W\,Hz$^{-1}$, \citep{Govoni2013}. 

In this work, we present new polarimetric observations of the galaxy cluster Abell 523 (hereafter A523). 
A523 consists of two subclusters that are undergoing a merger event along the SSW-NNE direction \citep{Girardi2016} and likely a secondary merger with a third subcluster along the ESE-WNW axis \citep{Cova2019}. The system hosts an extended and luminous diffuse synchrotron source with a luminosity at 1.4-GHz that is higher than expected from the typical correlation between radio and X-ray luminosity observed for radio halos \citep{Giovannini2011}. This emission shows bright polarized filaments at 1.4-GHz and a morphology that strongly differs from the distribution of the thermal gas, with a pronounced offset between the radio and X-rays peaks of about 310-kpc, in agreement with an intracluster magnetic field fluctating over large spatial scales of $\sim$1\,Mpc \citep{Girardi2016}.

\begin{figure*}
	\includegraphics[width=1.0\textwidth]{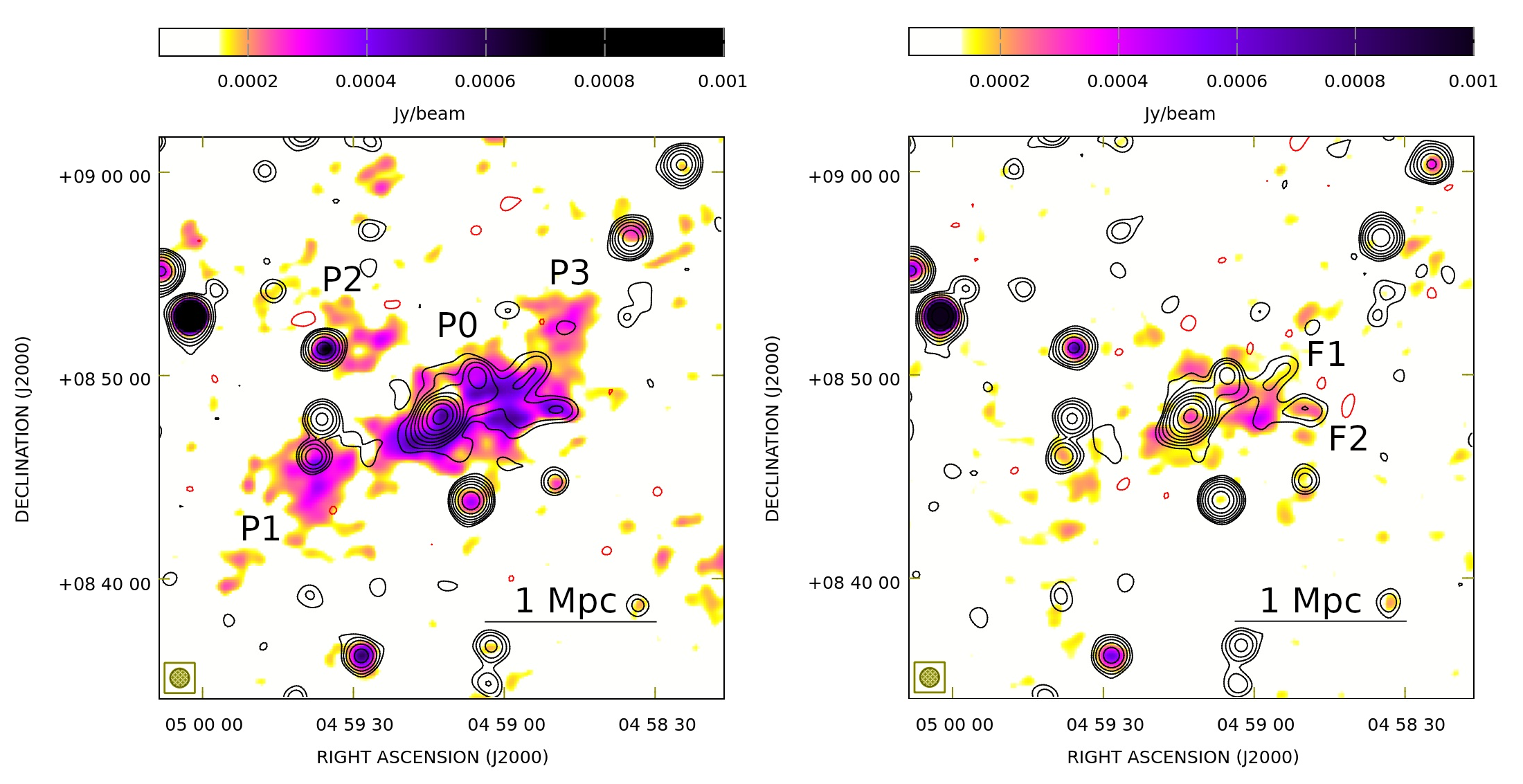}
    \caption{Emission in total intensity (contours) and polarization (colors) of the galaxy cluster A523 at 1.410-GHz (left) and 1.782-GHz (right). Black contours start at 3$\sigma_{\rm I}$ and scale by a factor 2, red contours are drawn at -3$\sigma_{\rm I}$. The synthesised beam is shown in the bottom left. In Table\,\ref{tab:B} we report the sensitivity values and resolution at the two frequencies. 
  }
    \label{fig1}
\end{figure*}

Recently, \cite{Vacca2022} found that the radio halo emission at 144-MHz of A523 is more extended than at 1.4-GHz up to a size of 1.8-Mpc, with an overall spectral index $\alpha_{\rm 144\,MHz}^{\rm 1.410\,GHz}\,=\,1.2\,\pm\,0.1$. By locally comparing the radio emission at 144-MHz and the X-ray emission in the energy band 0.5--2.5\,keV, they infer the presence of two components, one brighter in X-ray and dimmer in radio and the second the other way around, likely due to the complex dynamical state of the system. By investigating the properties of the system between 1.410-GHz and 1.782-GHz, \cite{Vacca2021} derived a steepening of the spectral index moving to higher frequencies with a mean value of $\alpha_{\rm 1.410\,GHz}^{\rm 1.782\,GHz}\,=1.8\pm0.7$, mainly corresponding to the filaments of the diffuse emission.

This paper is organized as follows. In \S\ref{radioobs} we present new Jansky Very Large Array (VLA) observations in polarization, in \S\ref{results} and \S\ref{discussion} we describe and discuss our results, and in \S\ref{conclusions} we present our conclusions. In the following, we adopt a $\Lambda$CDM cosmology with $H_{0}=67.4$\,km\,s$^{-1}$\,Mpc$^{-1}$, $\Omega_0=0.315$ and $\Omega_{\Lambda}=0.685$ \citep{Planck2020}. With this cosmology, at the distance of A523 ($z=0.104$, luminosity distance 499-Mpc), 1-arcsec corresponds to 1.98-kpc.

\section{Radio observations}
\label{radioobs}
We present VLA polarimetric observations in the frequency range 1--2\,GHz 
in D configuration obtained with a mosaic of six pointings (program 13A-168, PI M. Murgia). In Table\,\ref{tab:A} we summarize the observing setup and in Table\,\ref{tab:Aa} the coordinates of the pointing centres of the six pointings. The data were collected in full Stokes spectral line mode. The total initial bandwidth of 1-GHz consists of 16 spectral windows, 64-MHz and 64 channels each.  
The data were reduced following standard procedures using the NRAO's Astronomical Image Processing System (AIPS) package. Hanning smoothing was applied before bandpass calibration. The source J0319+4130 (3C84) was used as a bandpass and leakage calibrator, the source J0137+3309 (3C48) was used as a flux calibrator, the nearby source J0459+0229 was observed for complex gain calibration and finally the source J0521+1638 (3C138) to calibrate the polarization angle. The flux-scale by \cite{Perley2017} was adopted and the uncertainty in the flux density was assumed to be  2.5 percent \citep{Perley2013}.  Radio frequency interference (RFI) flagging has been applied by excision of data with values clearly exceeding the source flux. Following calibration, the data were spectrally averaged to 16 channels of 4-MHz per spectral window. After RFI excision, only the frequency ranges 1.315-1.508-GHz and 1.623-1.944-GHz survived for further inspection. Mosaic surface brightness images of these frequency bands were produced through single-scale Hogbom clean using the Common Astronomy Software Applications (\textsc{CASA}) package (task tclean) and including all the fields in Table\,\ref{tab:Aa}.
More details about the images presented in this paper are given in Table\,\ref{tab:B}.

\begin{table}
\caption{Setup of the VLA observations used in this work. Col.\,1, 2: central frequency and bandwidth; Col.\,3: VLA configuration; Col.\,4, 5: date and duration of the observations; Col.\,6: ID of the project.}
\centering
        \begin{tabular}{cccccc}
         \hline
          \hline
$\nu_{\rm c}$ & Bandwidth         &Config.      & Date & Duration&Project \\
 (GHz) & (MHz)&& & (min) &\\
\hline

 1.5& 1000& D& 2013 Jan 29&10.97& 13A-168\\
                \hline
            \hline
        \end{tabular}
        \label{tab:A}

\end{table}

\begin{table}
\caption{Coordinates of the pointing centres of the six VLA pointings.}
\centering
        \begin{tabular}{cc}
         \hline
          \hline
 RA & Dec \\
h:m:s&$^{\circ}$:$^{\prime}$:$^{\prime\prime}$\\
(J2000) &(J2000) \\
\hline

 04:59:40.4 & +09:01:59.35 \\
 04:58:39.6 & +09:01:59.35 \\
 05:00:10.7 & +08:48:59.70 \\
 04:59:10.0 & +08:49:00.00 \\
 04:58:09.3 & +08:48:59.70 \\
 04:59:40.3 & +08:36:00.50 \\
                  \hline
            \hline
        \end{tabular}
        \label{tab:Aa}

\end{table}
 
 \begin{table*}
\caption{Details of the images presented in this work. Col.\,1, 2: central frequency and bandwidth; Col.\,3: uv-range; Col.\,4: angular resolution; Col.\,5: maximum angular size accessible with the observations; Col.\,6, 7: sensitivity in Stokes I, and P respectively.}
        \begin{tabular}{ccccccc}
         \hline
          \hline
$\nu_{\rm c}$ & Bandwidth & uv-range&  Beam      & Max size & $\sigma_{\rm I}$& $\sigma_{\rm P}$\\
GHz               & MHz       &$\lambda$ & arcsec$\times$arcsec& arcmin    & mJy\,beam$^{-1}$& mJy\,beam$^{-1}$\\
\hline
1.410              &192       &158- 4835 &  56$\times$56    &  21.8  &0.15 & 0.05   \\
1.782              &320       &196- 6200 &  56$\times$56    &  17.5  &0.13 & 0.045     \\

\hline
\multicolumn{7}{c}{Rotation measure images}\\
\hline

1.362              &96        &158- 4835 &  56$\times$56    &  21.8  &0.15 & 0.06\\
1.458              &96        &158- 4835 &  56$\times$56    &  21.8  &0.15 & 0.06\\
1.782              &320       &196- 6200 &  56$\times$56    &  17.5  &0.13 & 0.045     \\
\hline
\hline
        \end{tabular}
        \label{tab:B}

\end{table*}

 \begin{figure*}
	\includegraphics[width=1.0\textwidth]{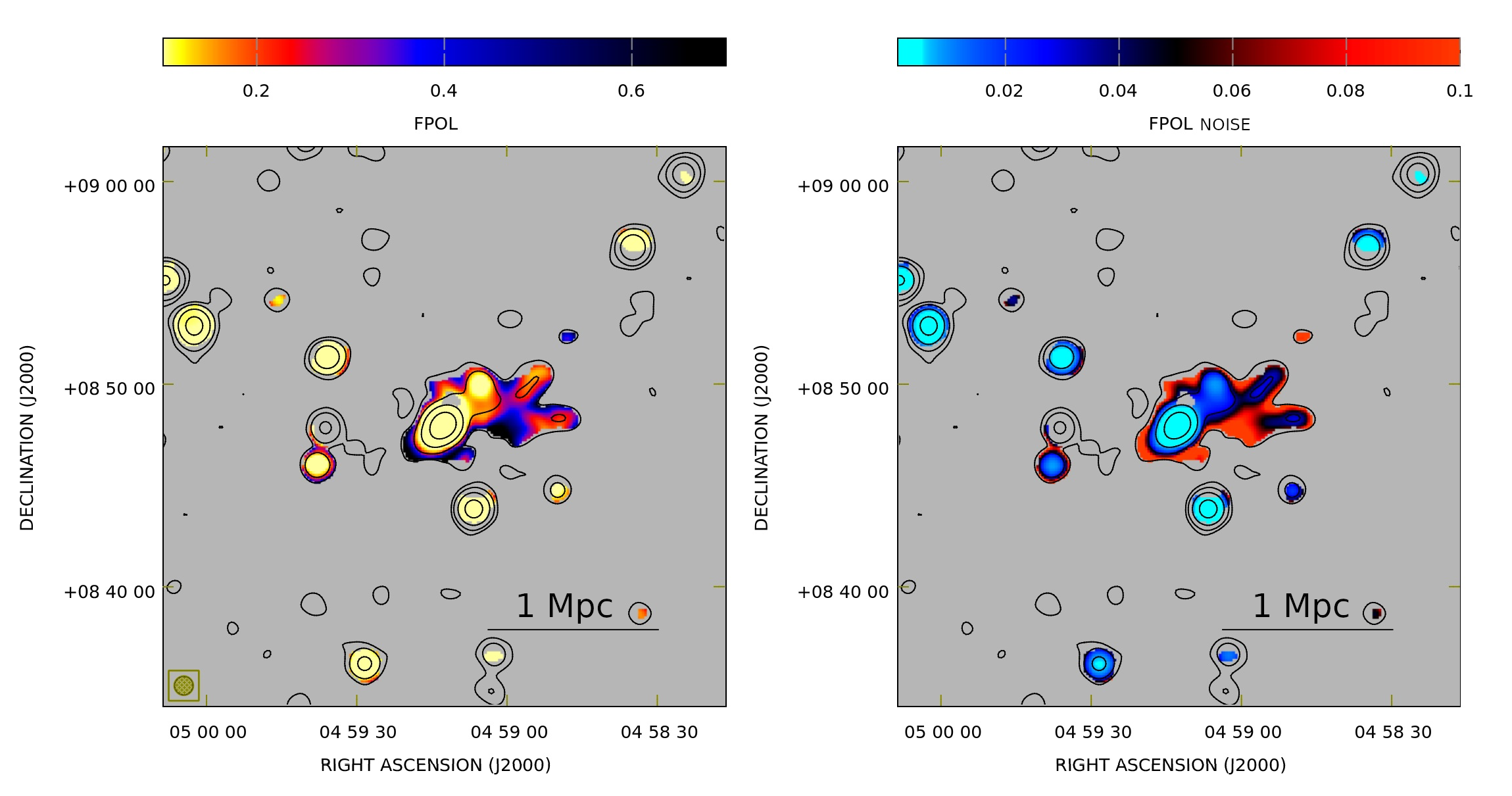}
    \caption{Fractional polarization (left) and its uncertainty (right) in the central region of the galaxy cluster A523 at 1.410-GHz after applying a 3$\sigma_{\rm FPOL}$ cut. Black contours start at 3$\sigma_{\rm I}$ and scale by a factor 4 in both panels. The synthesised beam is shown in the bottom left.  
  }
    \label{fig2a}
\end{figure*}

\begin{table*}
\caption{Properties of the polarization patches. Col.\,1: patch label (values for P0 are given without  applying any cut to the data and measuring the total intensity from the SRT+VLA image, see  \S\,\ref{Total intensity counterpart}); Col.\,2,3: coordinates of the polarization peak; Col.\,4: Mean total intensity brightness at 1.410-GHz; Col.\,5: Mean polarized brightness at 1.410-GHz; Col.\,6: Mean fractional polarization at 1.410-GHz; Col.\,7: Mean total intensity brightness at 144-MHz (see section \S\,\ref{results} for more details). The boxes used for the statistics are shown in Appendix\,\ref{appendix_c}.}
\centering
        \begin{tabular}{cccccccc}
         \hline
          \hline
 Patch &RA & Dec & I$_{\rm 1.410\,GHz}$ &P$_{\rm 1.410\,GHz}$ & FPOL$_{\rm 1.410\,GHz}$ & I$_{\rm 144\,MHz}$\\
&h:m:s&$^{\circ}$:$^{\prime}$:$^{\prime\prime}$& mJy\,beam$^{-1}$ & mJy\,beam$^{-1}$ &  percent&mJy\,beam$^{-1}$\\
&(J2000) &(J2000) \\
\hline
P0& 04:58:58.4 & +08:47:52.27 &0.93&0.22&24&22.8 \\
P1& 04:59:37.1 & +08:44:29.18 &<0.45&0.21&>47&2.7\\
P2& 04:59:24.3 & +08:51:50.78 &<0.45&0.17&>37&1.8\\
P3& 04:58:47.0 & +08:52:39.10 &<0.45&0.21&>47& 3.1\\
 
                  \hline
            \hline
        \end{tabular}
        \label{tab:C}

\end{table*}

\section{Results}
\label{results} 

In Fig.\,\ref{fig1} we present the total intensity and polarized emission (corrected for the
positive bias) in the central region of the galaxy cluster A523 at 1.410-GHz (left panel) and at 1.782-GHz (right panel). Contours describe the total intensity emission, while colors represent the polarized emission above 3$\sigma_{\rm P}$\footnote{ $\sigma_{\rm I}$ and $\sigma_{\rm P}$ indicate the noise rms of the total intensity and polarization images respectively.}. 
The theoretical thermal noise expected at 1.410-GHz (see Table\,\ref{tab:A} and Table\,\ref{tab:B} for details about the observing setup used for the images) is $\approx\,0.037\,$mJy\,beam$^{-1}$. However, at this frequency and at our spatial resolution, Stokes I images are dominated by the confusion noise, being $\approx\,0.13-0.15\,$mJy\,beam$^{-1}$ \citep{Condon2002,Loi2019a} after a few minutes of observation, while Stokes Q and U are not. These values are comparable with the noise in our images.  

These images reveal four patches of polarized emission: P0 associated with the total intensity signal and already known from \cite{Girardi2016}, as well as three new patches extending beyond the total intensity signal, the brightest one in the south east of the cluster (P1), a second one in the north east (P2) and a third one starting at the location of the two total intensity filaments  F1 and F2 and extending to the north (P3, see labels in Fig.\,\ref{fig1}). 
To confirm the existence of these  new patches of polarized emission, we re-examined the data by \cite{Girardi2016} who applied a 3$\sigma$ cut in fractional polarization in the analysis. Differently from them, we do not apply any cut to the data aiming at  detecting all the polarized emission present in the field. Our re-analysis confirms the presence of the large-scale polarized signal also in these archival data, see Appendix\,\ref{appendix_a}. As described in the appendix this signal appears to be real and not an artifact.

When comparing the images at 1.410-GHz and 1.782-GHz, we note that both the total intensity and the polarized emission rapidly fades at high frequency and only the brightest regions in polarization survive mostly coinciding with the total intensity filaments F1 and F2.
The overall size of the source at 1.410-GHz is 21-arcmin, i.e. about 2.5-Mpc at the distance of the cluster. Integrating the polarized intensity down to 
$3\sigma_{\rm P}$, we derive a polarized flux density of $25.7\pm0.8$\,mJy at 1.410-GHz and of $9.4\pm0.4$\,mJy at 1.782-GHz. We masked the compact sources embedded in the diffuse emission and assumed at their location a brightness of the diffuse emission equal to the average brightness over the full source.

The data presented here have been corrected for the on-axis but not for off-axis instrumental polarization. In order to rule out the possibility that this emission is due to residual instrumental off-axis polarization, we measure the peak in total intensity radio brightness of the sources with peak above 30$\sigma_{\rm I}$ at 1.410-GHz in the full field of view. Their sum amounts to $\approx\,$210\,mJy. 
 Since the instrumental off-axis polarization for VLA snapshot observations should be at most $\sim$\,2 percent \citep{Cotton1994} of the total intensity peak if not corrected,

such a contribution would be at most of $\approx\,4\,$mJy at 1.410-GHz in our case, suggesting that this emission can not be explained by this phenomenon. 

In the following, we describe the polarization properties of the polarized emission associated with the diffuse emission in A523 as well as the new patches presented in this paper. A summary is given in Table\,\ref{tab:C}.

\subsection{Central emission}
\label{Central emission}
In order to compare our results with those available in the literature, we analyze first the properties of the patch P0. 
The fractional polarization distribution and its uncertainty at 1.410-GHz in the central region of the cluster are shown in Fig.\,\ref{fig2a}, after applying a cut of 3$\sigma_{\rm FPOL}$. Using this image, we derive the mean fractional polarization of (33$\pm$8) percent at 1.410-GHz .  Similarly, we derive the mean fractional polarization after applying a 3$\sigma_{\rm FPOL}$ cut at 1.782-GHz and obtain (34$\pm$9) percent. %at 1.782-GHz. 
\cite{Girardi2016} estimate the average fractional polarization at 1.4-GHz with concentric annuli centred on the radio centroid and find a rather flat profile at $\approx$ 15–20 per cent (see their fig. 19). Their value is slightly lower than  ours, likely because
they recover larger angular scales in total intensity due to the better uv-coverage (see the following and Appendix\,\ref{appendix_a}). 

 In Fig.\,\ref{fig2}, we show the distribution of the polarization angle orientation with black vectors, whose length is proportional to the polarized intensity, without any cut in fractional polarization. The vectors are overlaid on the total intensity and polarized emission in the galaxy cluster A523 at 1.410-GHz respectively in black contours and colors.
The blue contours represent the X-ray emission as collected from the three EPIC instruments on board of \textit{XMM-Newton} in the soft 0.5--2.5\,keV band for a total exposure time of about 220-ks by \cite{Cova2019}. 
The spatial distribution of the polarization angle in Fig.\,\ref{fig2} for the central region is very similar to that shown by \cite{Girardi2016}, see fig. 17 in their paper, with peaks of fractional polarization $\gtrsim$50 per cent.

\subsection{New polarized patches}
\label{New polarized patches}
\begin{figure}
		\includegraphics[width=0.5\textwidth]{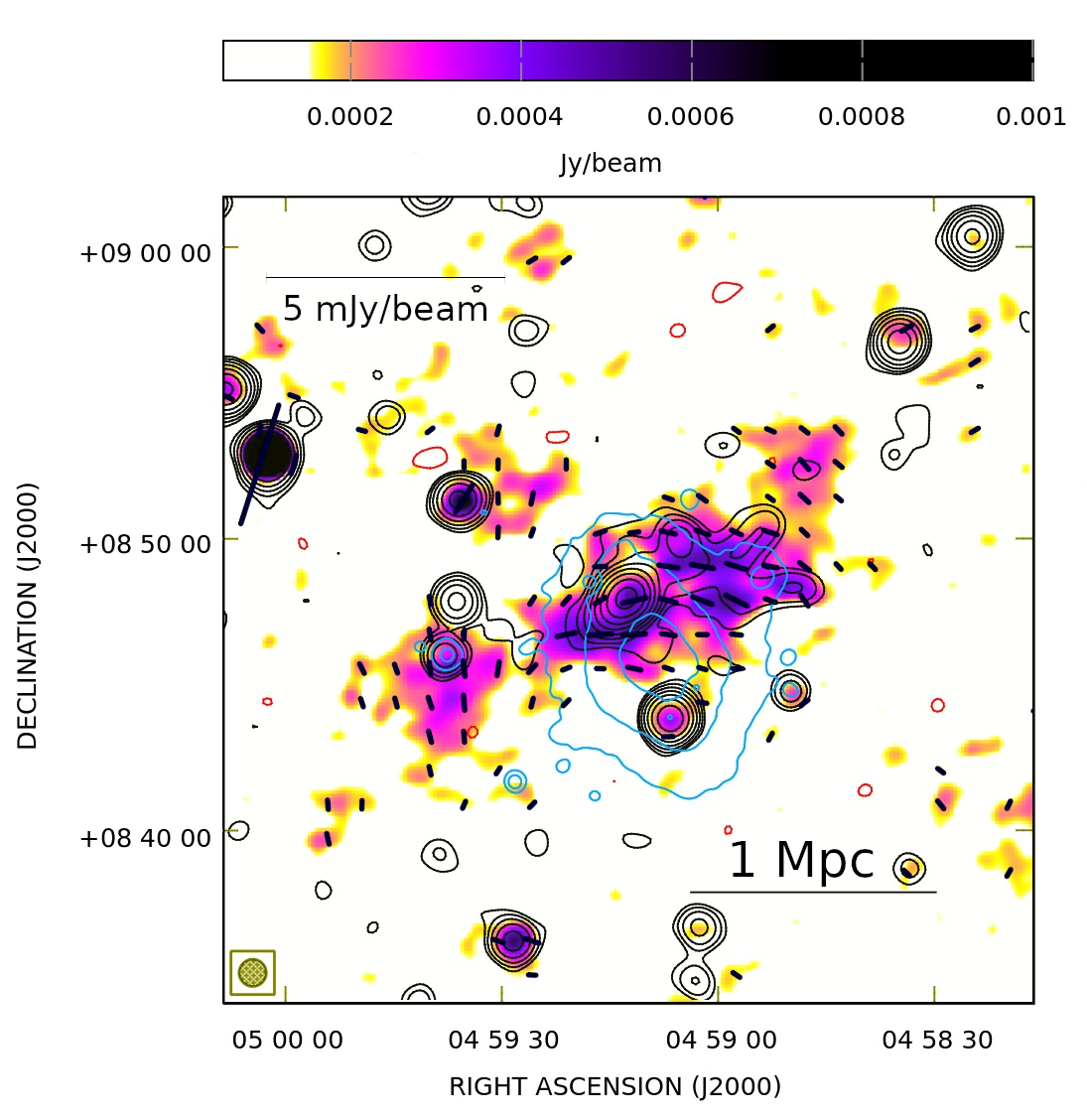}
    \caption{Polarized (colors) and total intensity (contours) radio emission from the galaxy cluster A523 at 1.410-GHz: black contours start at 3$\sigma_{\rm I}$ (see Table\,\ref{tab:B}) and scale by a factor 2, while red contours are drawn at -3$\sigma_{\rm I}$. Vectors represent the polarization angle orientation, while their length is proportional to the polarized intensity (see horizontal bar in the top left corner). Blue contours show the X-ray emission as observed with \textit{XMM-Newton} \citep[][see text for more details]{Cova2019}, starting from 10\,counts\,s$^{-1}$\,deg$^{-2}$ and scaling by a factor 2. 
    }
    \label{fig2}
\end{figure}

The fractional polarization of P1, P2, and P3 is not visible in Fig.\,\ref{fig2a} because of the cut at 3$\sigma_{\rm FPOL}$, while their polarized brightness and polarization angle distribution can be clearly observed in Fig.\,\ref{fig2}. 
These new patches of polarized emission are located at the boundary of the X-ray emission of the cluster, where the thermal gas density associated with the system is expected to be lower and the magnetic field to be weaker. By drawing a box at the location of these patches (see Appendix\,\ref{appendix_c}), we measure a mean polarized brightness respectively of 0.21, 0.17, and 0.21\,mJy\,beam$^{-1}$ at 1.410-GHz. Considering an upper limit in total intensity of 3$\sigma_{\rm I}=0.45$\,mJy\,beam$^{-1}$, we derive a lower limit for the mean fractional polarization respectively of $FPOL=P/I>$47  percent, 37 percent, and 47 percent. At 1.782-GHz both the polarized and the total intensity brightness are below 3$\sigma$.

\begin{figure*}
		\includegraphics[width=0.9\textwidth]{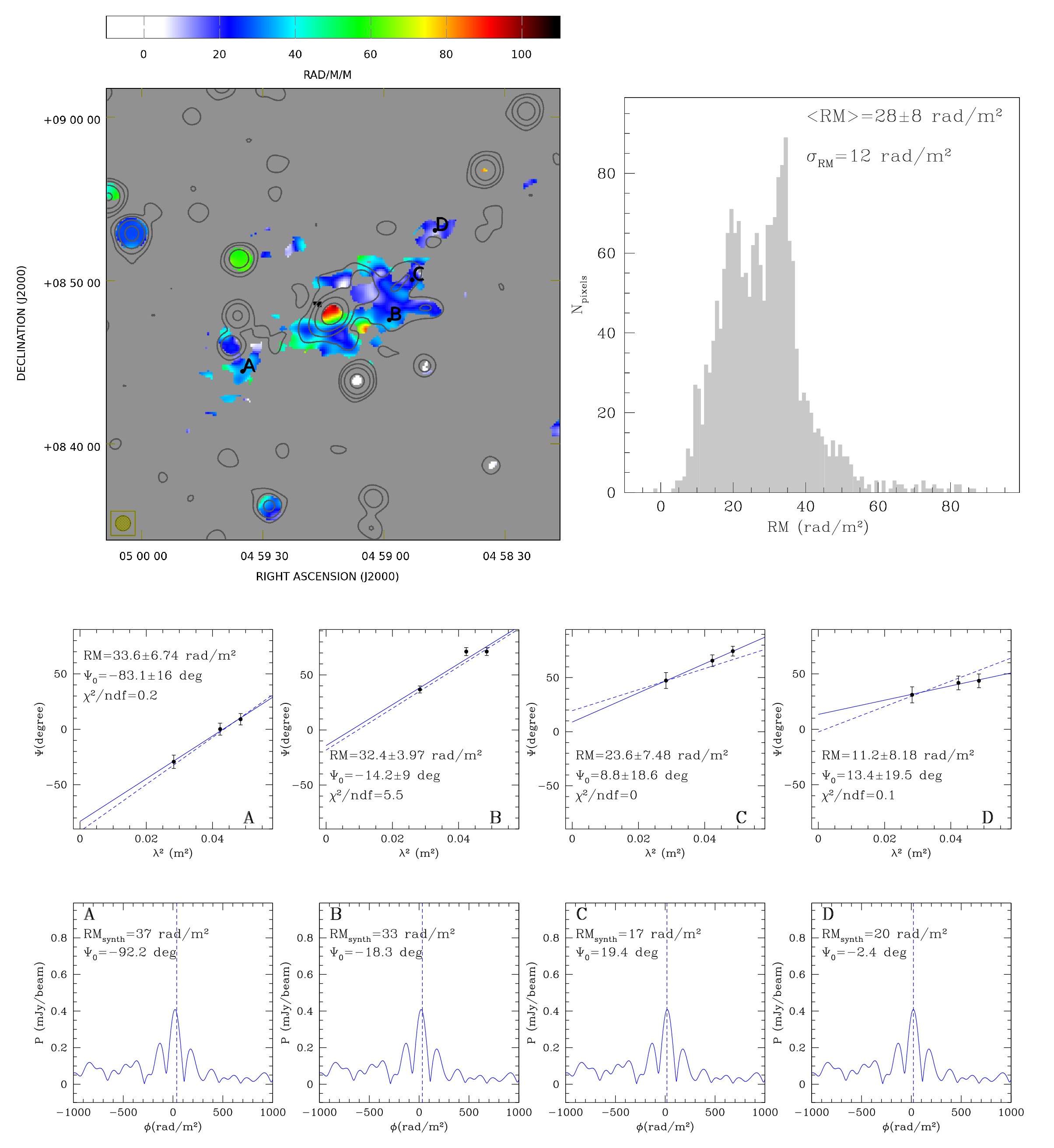}
    \caption{Top left panel: Rotation measure image of the polarized emission in A523. Top right panel: histogram of the RM after masking the embedded discrete sources. Middle panels: Polarization angle versus $\lambda^2$ for pixels at different locations of the diffuse emission (A, B, C, and D). The solid blue line describes the linear fit, 
    while the dashed blue line has a slope corresponding to the Faraday depth and an intrinsic polarization angle derived through the RM synthesis technique. Bottom panels: Faraday spectra derived at the location of pixels A, B, C, and D. The dashed line is drawn at the Faraday depth corresponding to the peak in polarized intensity. See text for more details. }
    \label{fig3}
\end{figure*}
\begin{figure}
	\includegraphics[width=0.5\textwidth]{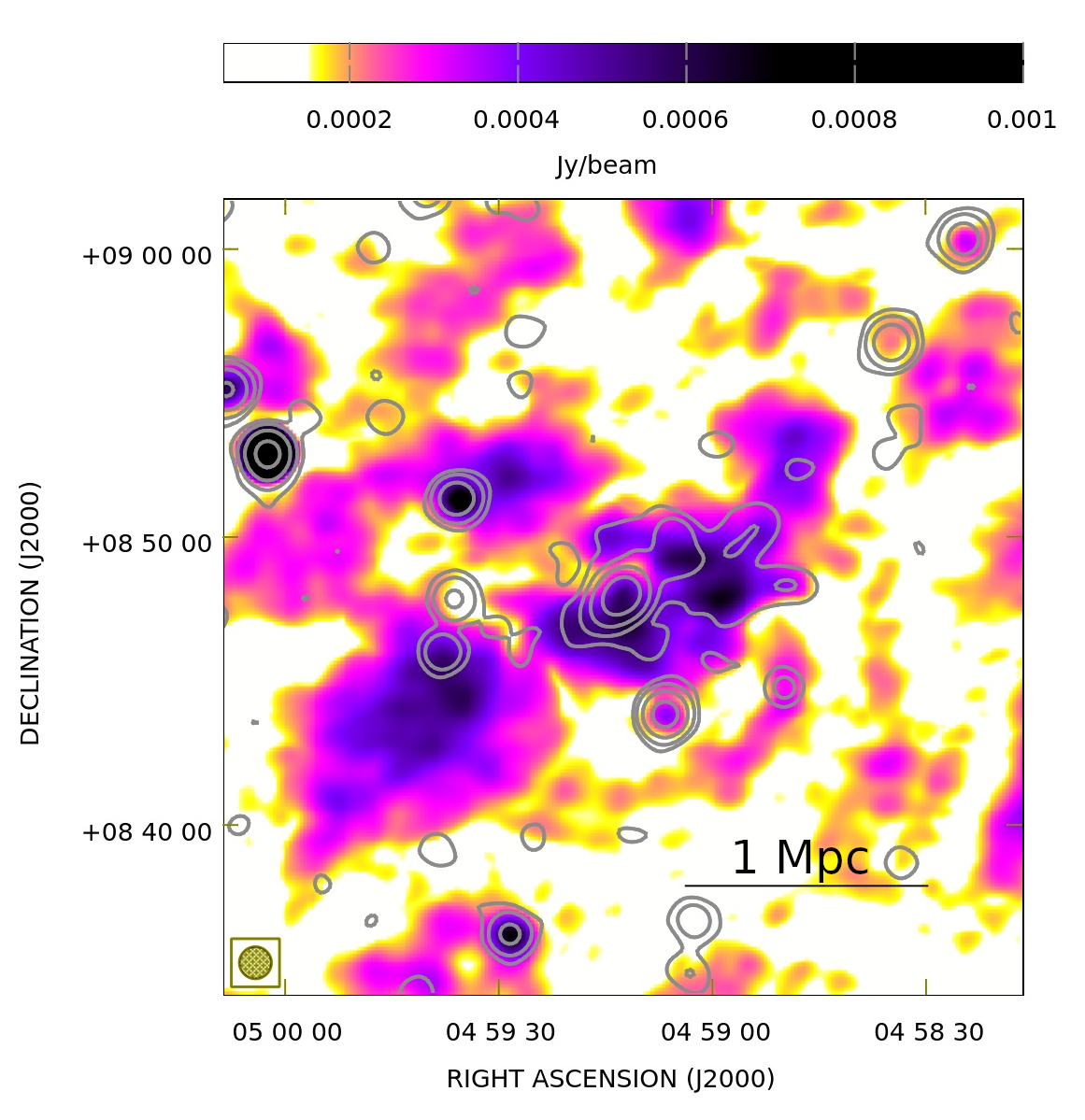}
    \caption{Polarized radio emission from the galaxy cluster A523 corresponding to the peak in the Faraday spectrum in colors overlaid on gray contours that represent the total intensity at 1.410-GHz. The contours start at 3$\sigma_{\rm I}$ (see Table\,\ref{tab:B}) and scale by a factor 4.}
    \label{fig4a}
\end{figure}
\begin{figure*}
	\includegraphics[width=\textwidth]{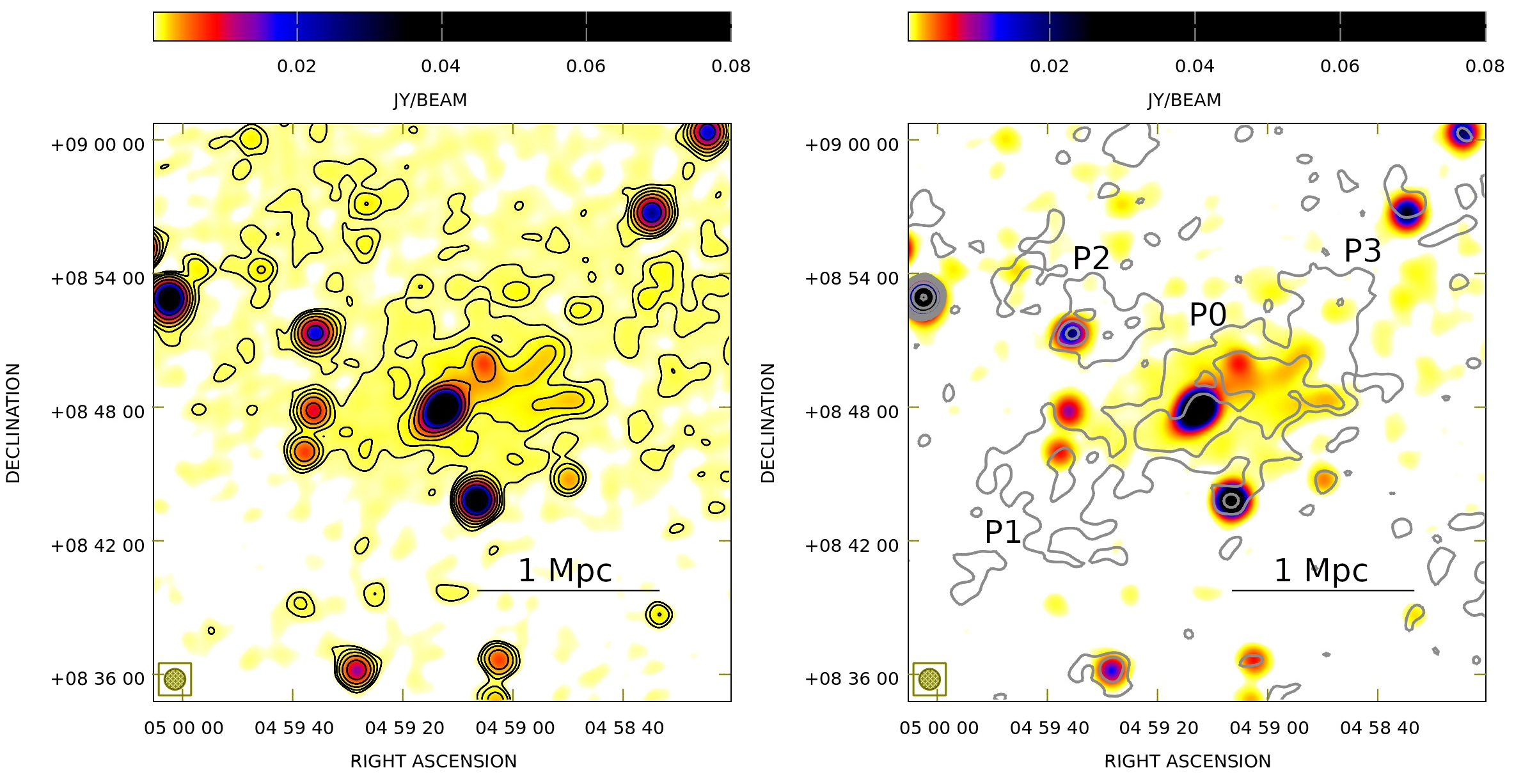}
    \caption{Left panel: total intensity SRT+VLA image in contours and colours. Contours start at 3$\sigma=$4.5\,mJy\,beam$^{-1}$ and scale by a factor 2, colors show the emission above 1$\sigma$. Right panel: total intensity SRT+VLA image in colours above 3$\sigma$. Contours represent the polarized emission at 1.410-GHz from VLA data. They are drawn at 3$\sigma_{\rm P}=$0.15\,mJy\,beam$^{-1}$ and scale by a factor 2. The synthetized beam is shown at the bottom left corner. }
    \label{fig5}
\end{figure*}
\begin{figure*}
	\includegraphics[width=\textwidth]{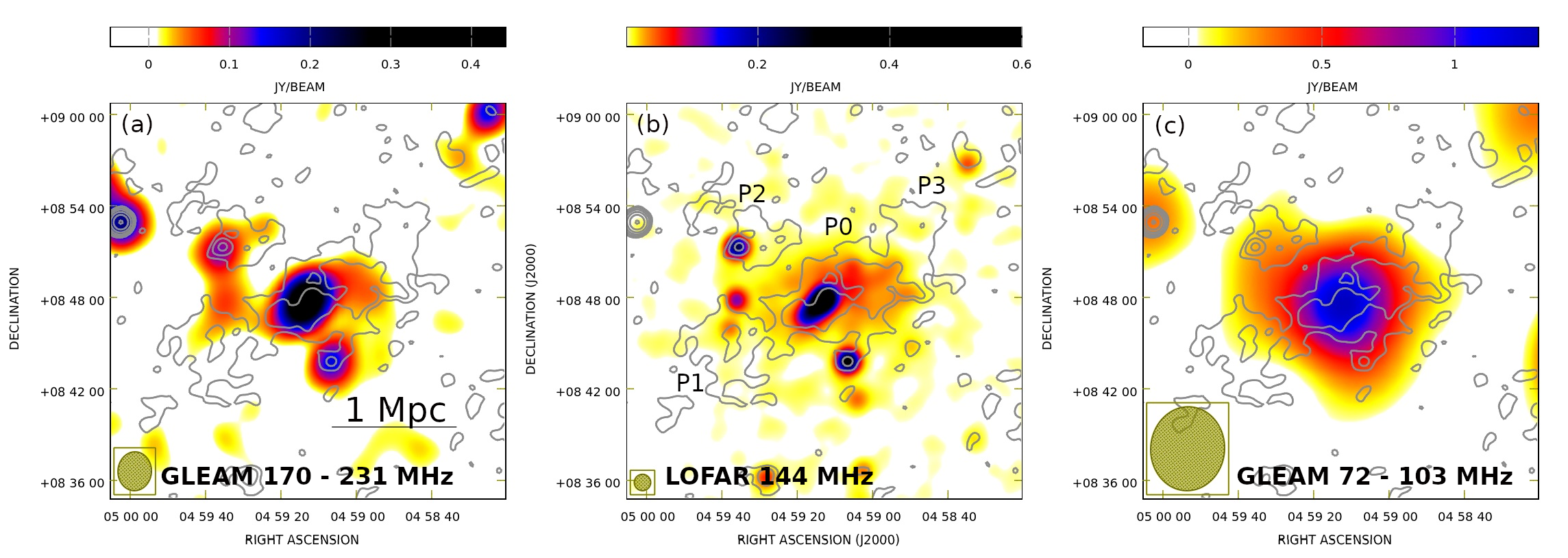}
    \caption{Contours at 3$\sigma_{\rm P}$ of the polarized emission detected at 1.410-GHz are superimposed on 
    the GaLactic and Extragalactic All-sky MWA Survey (GLEAM) image in the frequency range 170--231\,MHz (\protect\citealt{Wayth2015,Hurley-Walker2017}, 153.5\,arcsec$\times$ 132.4\,arcsec, panel a), the LOFAR image in the frequency range 120--168\,MHz (\protect\citealt{Vacca2022}, 65\,arcsec$\times$65\,arcsec, panel b), and the GLEAM image in the frequency range 72--103\,MHz (\protect\citealt{Wayth2015,Hurley-Walker2017}, 330.9\,arcsec$\times$ 291.4\,arcsec, panel c). Images in colors show all the emission above their 1$\sigma$ (respectively 11\,mJy\,beam$^{-1}$, 1.3\,mJy\,beam$^{-1}$, and 30\,mJy\,beam$^{-1}$).}
    \label{fig5a}
\end{figure*}

\subsection{Rotation Measure analysis}
\label{Rotation Measure analysis}

In order to mitigate bandwidth depolarization and better investigate how the polarization properties vary with frequency, we divide the full frequency range in three sub-bands: the first two 96-MHz wide centred respectively at 1.362 and 1.458-GHz and the third located at 1.782-GHz and with a bandwidth of 320-MHz\footnote{At high frequency the bandwidth depolarization is less important  and a larger bandwidth ensures to detect enough signal.}, see Table\,\ref{tab:B}. 
In the top left panel of Fig.\,\ref{fig3}, we present the RM image of the  cluster, produced with the software \textsc{FARADAY} \citep{Murgia2004}. Using as input the Q and U images for the three frequency sub-bands, the algorithm fits pixel by pixel the observed polarization angle $\Psi$ versus the squared observing wavelength $\lambda^2$. A few iterations have been performed in order to reduce n$\pi$ ambiguity issues, starting from a subset of high signal-to-noise pixels and gradually including lower signal-to-noise pixels. We utilized only those regions of the source with a polarized signal above 3$\sigma_{\rm P}$ in the lower frequency sub-band and pixels with an uncertainty in polarization angle below 10$^{\circ}$ at all the three frequencies. 

The RM image shows a uniform distribution over the full polarized emission with a mean value of $\langle$ RM $\rangle=(28\pm8)\,$rad\,m$^{-2}$ and a dispersion $\sigma_{\rm RM}=12\,$rad\,m$^{-2}$, after excluding embedded discrete sources. A histogram of the overall distribution is shown in the top right panel of Fig.\,\ref{fig3}. In the RM image labels A, B, C and D are used to identify 
four pixels at different locations of the diffuse emission. At the location of these pixels the observed polarization angle $\Psi$ versus the squared observing wavelength $\lambda^2$ is shown in the middle plots as well as the best-fitting of the trend. The plots for A, C, and D show a linear trend, indicating that the rotation of the polarization angle is likely due to a completely foreground medium. For point B, the closest in projection to the cluster centre, the behaviour slightly differs from a linear one, possibly indicating a partial mixing between the emitting radio plasma and the magneto-ionic medium.

In order to better investigate if there is any internal depolarization missed by the $\lambda^2$-fitting, we applied the RM synthesis technique \citep{Brentjens2005} to the data including all the frequency range available (i.e., both subbands at 1.410-GHz and 1.782-GHz). We produced Q and U cubes 
using channels 4-MHz wide to cover the full frequency range available with a spatial resolution of 67-arcsec imposed by the low frequency channels. We applied the RM synthesis spanning from -1000\,rad\,m$^{-2}$ to 1000\,rad\,m$^{-2}$ with a step of 1\,rad\,m$^{-2}$ and a full width half maximum of 100\,rad\,m$^{-2}$. In the bottom panels of Fig.\,\ref{fig3}, we show the Faraday spectrum at the location of pixels A, B, C and D. At our resolution in Faraday depth a single peak is clearly visible, but we note that we did not apply any clean in Faraday space. At these locations and overall in the rest of the image, the Faraday depth values obtained with the RM synthesis approach show a good agreement with the RM values from $\lambda^2$-fitting, suggesting that the rotation of the polarization angle of the diffuse emission is mainly due to a foreground medium and internal depolarization can be neglected. 

In Fig.\,\ref{fig4a} we show the peak in polarization obtained with the RM synthesis approach overlaid with the total intensity contours at 1.410-GHz. 
Integrating the polarized intensity in the same region as before (above the $3\sigma_{\rm P}$ at 1.410-GHz), we derive a polarized flux density\footnote{The flux density value has been obtained after correction for the primary beam response.} of $29.1\pm0.8$\,mJy. As before, we masked the compact sources embedded in the diffuse emission and assumed at their location a brightness of the diffuse emission equal to the average brightness over the full source. The polarized flux density obtained with the RM synthesis technique is about 13 percent larger than the value measured from Fig.\,\ref{fig1} (left panel), as expected in case of in-band depolarization. We also note that here we are integrating over the full bandwidth available, while the left panel in Fig.\,\ref{fig1} refers to a smaller portion of the frequency range. % considered here is larger than in \S\,\ref{results}.
A few additional patches of polarized brightness comparable to P0, P1, P2 and P3 emerge at large distance from the cluster centre. Since these patches are located in regions of the image where the primary beam correction is $\gtrsim 70$ percent, a more detailed analysis is needed to understand if they are real, but is left for future work.

According to \cite{Hutschenreuter2021}, the mean Galactic Faraday rotation in the direction of this galaxy cluster is $\langle$RM$_{\rm gal}\rangle\,\approx$\,38.8\,rad\,m$^{-2}$ with an uncertainty of $\sigma_{\rm RM, gal}\,\approx\,$21.3\,rad\,m$^{-2}$ within a distance of 5$^{\circ}$ from the cluster centre, assumed to be at RA 04h:59m:06.2\,s and Dec +08$^{\circ}$:46$^{\prime}$:49$^{\prime\prime}$. 
The $\langle$RM$\rangle$ derived from our analysis and the $\langle$RM$_{\rm gal}\rangle$ are consistent within the uncertainties.

\subsection{Total intensity counterpart}
\label{Total intensity counterpart}

The observations presented in this work indicate the presence of a large-scale polarized radio emission without a clear total intensity counterpart. This effect is due to two observational limitations that we can describe as follows:\\

\noindent
1) As shown in Table\,\ref{tab:B}, the noise in polarization $\sigma_{\rm P}$ of our images is about three times lower than in total intensity $\sigma_{\rm I}$. While low-resolution radio observations in total intensity are limited by the confusion noise from blending of background radio sources, polarized data are not because of the lower number-density of polarized sources and can therefore reach higher sensitivities \citep{Loi2019b}. By comparing our different sensitivities in total intensity and polarization, we derive that we can detect faint sources with a fractional polarization larger than 30 percent.\\

\noindent
2) Another limitation that can have a significant impact on total intensity imaging of large-scale emission is represented by the minimum baseline of interferometers. This baseline sets a limit to the largest angular scale that can be sampled by the interferometer, preventing reconstruction of the full size of large-scale diffuse synchrotron sources in total intensity. 
The largest angular scale structure that can be recovered\footnote{Source \url{https://science.nrao.edu/facilities/vla/docs/manuals/oss/performance/resolution}} with full synthesis VLA observations at 1.5-GHz in D configuration is about 16-arcmin (i.e., 1.9-Mpc). This number should be divided by two in case of single snapshot observations. The observations presented in this work consist of six pointings 10.97\,min each, for a total of 1.2\,h of observation, therefore more comparable with a single snapshot observation than a full synthesis one. Indeed, in total intensity we recover a size of the diffuse emission of about 8.5-arcmin (i.e., 1-Mpc), consistent with the largest-angular scale accessible with short exposure observations, while longer exposure data allow a better reconstruction of the source (see, e.g., Fig.\,\ref{figA}).

Due to the variation of the polarization angle of the electric-vector, either intrinsic or due to the Faraday rotation, 
the polarization angle transfers power to smaller scales with respect to scales characterizing the total intensity signal \citep[see, e.g., ][]{Bernardi2003}. As a consequence, the
polarized emission is expected to show a patchy structure that enables its imaging also when associated with large-scale radio sources.
While observing at 325-MHz and 610-MHz the Galactic disk, \cite{Wieringa1993} revealed the presence of structures in Q and U, not observed in total intensity. They explained the absence of Stokes I emission as a result of missing short spacing that do no affect the polarized intensity because polarization shows structures on smaller scales, as confirmed later by \cite{Haverkorn2002}. We are likely observing a similar phenomenon here. By inspecting Fig.\,\ref{fig2}, indeed the different polarization patches show a different orientation of the polarization angle, uniform within the patch itself, suggesting a fluctuation scale of the polarization signal comparable to the size of the patches. Other cases can be found in literature. \cite{Bernardi2003} indeed present observations at 1.4-GHz of the BOOMERanG field, a high Galactic latitude field, where they find evidence of a smooth bright polarized Galactic emission with no corresponding feature in total intensity.

Observations performed with single-dishes can reconstruct total intensity emission up to scales comparable with the size of the scan performed on the sky. Combination of single-dish data with interferometric observations over the same area and in the same frequency range offers the possibility to recover large-scale emission discriminating at the same time possible embedded compact sources, down to the sensitivity of the datasets.
Therefore, we combined 
the interferometric data presented in this work with single-dish SRT data by \cite{Vacca2018}. The resulting image is shown in Fig.\,\ref{fig5}. 
The data have been combined after selecting the same frequency range and field of view and after aligning the SRT flux scale with the NVSS, see section 4 in \cite{Vacca2018} for details about the procedure.
The image shows a better reconstruction of the total intensity emission associated with the patch P0, comparable with the size of the polarized emission of this patch and with total intensity images from archival data characterized by longer exposure time (see Appendix\,\ref{appendix_a}). If we measure the total intensity from the SRT+VLA image in order to not miss large scale emission because of the short spacing issue and do not apply any cut to the data in fractional polarization, we 
measure at 1.410-GHz a fractional polarization of 24 percent, {value that is} better agreement with results by \cite{Girardi2016}. 
No clear indication of a total intensity counterpart is instead detected at the location of P1, P2 and P3 even in the combined SRT+VLA image, suggesting that this emission is below the noise level of our observations.

Low frequency observations are better suited to sampling large-scales (e.g., up to $\approx$\,1\,$^{\circ}$ in the case of the LOw Frequency ARray, LOFAR, at 144-MHz) as well as
for detecting steep spectrum sources, as radio halos are difficult to
observe at frequencies $\gtrsim$\,1\,GHz. Therefore,  
we examined the presence of total intensity counterparts in images obtained from interferometers operating at lower frequencies. In Fig.\,\ref{fig5a}, we show the emission at the centre of the galaxy cluster A523 as observed from 231-MHz down to 72-MHz by LOFAR \citep{Vacca2022} and the Murchison Widefield Array \citep[MWA, ][]{Wayth2015,Hurley-Walker2017}.
At the location of the polarization patches hints of a total intensity signal are present especially at 144-MHz with LOFAR data (central panel of Fig.\,\ref{fig5a}). By using the same box as in \S\,\ref{New polarized patches} at the location of P1 in the LOFAR image, we measure a mean radio brightness of 2.7\,mJy\,beam$^{-1}$ and a total flux density of $36.3\,\pm\,8.7$\,mJy. In order to assess the significance of this measurement, we placed eight boxes with the same size randomly in the image noise and estimated the mean radio brightness and its standard deviation. The mean of the means is -0.26\,mJy\,beam$^{-1}$ and is characterized by a standard deviation of 0.18\,mJy\,beam$^{-1}$, indicating that the signal detected at the location of P1 is significant. By inspecting the LOFAR image at higher resolution (9-arcsec, \citealt{Vacca2022}), no compact source is identified at the location of the LOFAR radio brightness peak in the region corresponding to P1 at the sensitivity and resolution of the available data. For P2 we measure a mean radio brightness of 1.8\,mJy\,beam$^{-1}$ and a total flux density of $14.5\,\pm\,4.7$\,mJy, and for P3 a mean radio brightness of 3.1\,mJy\,beam$^{-1}$ and a total flux density of $32.9\,\pm\,7.8$\,mJy. These measurements indicate that a brightness excess is present also at the locations of these patches.

\begin{figure*}
		\includegraphics[width=0.9\textwidth]{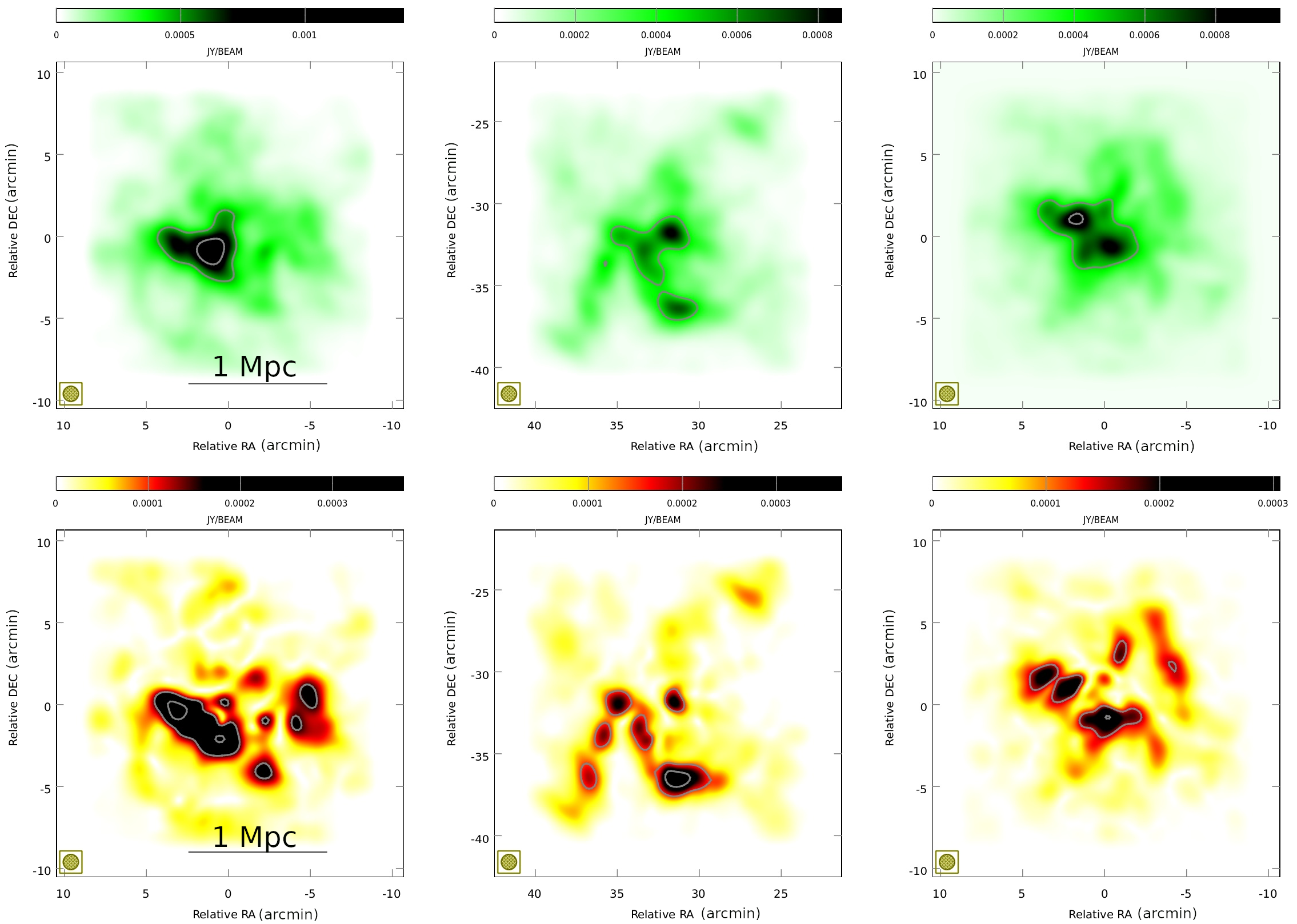}
   \caption{Simulated emission in colors and contours in total intensity (top panels) and polarization (bottom panels) of a galaxy cluster at a central frequency of 1.410-GHz with a bandwidth 192-MHz corresponding to different viewing angles (FWHM=56\,arcsec). In the top panels, contours start at 0.45\,mJy\,beam$^{-1}$, i.e. the 3$\sigma_{\rm I}$ at 1.410-GHz, and scale by a factor 2. In the bottom panels,  contours start at 0.15\,mJy\,beam$^{-1}$, i.e. the 3$\sigma_{\rm P}$ at 1.410-GHz, and scale by a factor 2. The synthesised beam is shown in the bottom left of each image.}
   \label{fig6}

\end{figure*}
\section{Discussion}
\label{discussion} 
%In conclusion t
This is the first time that a polarized signal is observed at the centre of a galaxy cluster on scales larger than 1-Mpc. 
This polarized emission consists of multiple patches that cover a full size of $\approx\,21$\,arcmin (i.e., 2.5-Mpc). We are blind to the total intensity counterpart of this emission at 1.410-GHz possibly 
due to the sensitivity of the images, limited by the confusion noise. 
Overall, the analysis of the polarimetric properties of the source indicates that
\begin{itemize}
    \item the trend of the polarization angle $\Psi$ versus the observing wavelength squared $\lambda^2$ is linear;
    \item the Faraday depth derived through the rotation measure synthesis technique is consistent within the errors with the RM derived from the $\lambda^2$-fitting technique;
    \item the Galactic RM is comparable within the errors with the mean RM of the source;
    \item the new polarized patches discovered are located in the outskirts of the X-ray emission of the galaxy cluster.
\end{itemize}
This polarized emission is difficult to interpret and in the following we discuss possible scenarios.

\subsubsection*{Galactic structure}
The galaxy cluster is located at low Galactic latitude (b=-20.15$^{\circ}$). This could suggest that the source of the polarized emission is a Galactic structure. The possibility of a distinct Galactic object is disfavoured, since when inspecting this region of the sky with all-sky synchrotron and thermal bremsstrahlung images available in the literature, no hint of a possible Galactic source emerges, see Appendix\,\ref{appendix_b} for more details. The images used here are characterized by a low spatial resolution (between 6-arcmin and 51-arcmin). Higher resolution images would be necessary in order to completely exclude the presence of smaller-scale Galactic objects as, e.g., supernovae remnants. An alternative possibility is that we are seeing a polarization structure caused by a feature of the Galactic magnetic field. Polarization observations over a larger field of view would help to assess the likelihood of this hypothesis.

\subsubsection*{Cluster radio relic}
The peripheral location of the patches P1, P2 and P3 and their high polarization could suggest a radio relic interpretation, 
 since radio relics are typically observed
%since typically radio relic are observed 
at the outskirt of galaxy clusters and are characterized by fractional polarization larger than 20 percent at 1.4-GHz \citep[see, e.g.,][]{vanWeeren2019}. 

According to the results from optical observations the direction of the primary merger is SSW-NNE with only a small line-of-sight velocity difference between the north and south subclusters, indicating that the merger axis is almost perpendicular to the line of sight
\citep{Girardi2016,Golovich2019}. The X-ray analysis by \cite{Cova2019} suggests a secondary merger process, but still on the plane of the sky. Considering this scenario, we could expect a shock wave propagating on the plane of the sky that would cause a polarized radio relic via synchrotron emission from shock accelerated electrons in shock compressed magentic fields \citep{Ensslin1998}, but no detection is available in literature to date. A merger component along the line of sight could be missed if the merger is at the turnaround point or in case the secondary merger tentatively detected by \cite{Cova2019} is not entirely in the plane of the sky. The detection of such a shock wave could be hindered due to projection effects and therefore missed by present X-rays observations.

From the point of view of the radio data, at 1.4-GHz the bulk of the emission sits in the north of the system \citep{Giovannini2011}. Observations at 144-MHz reveal a more extended source characterized by a complex morphology
consisting of two bright filaments in the north and a further extension towards the south consisting of an additional filament and a roundish patch of emission \citep[see ][]{Vacca2022} following the optical overdensity and the X-rays elongation presented respectively by \cite{Girardi2016} and \cite{Cova2019}. On the basis of the multi-frequency data presently available, as already discussed in \cite{Vacca2022}, a possibility is that we are looking at the superposition of different sources. The radio morphology and the spectral index distribution could support a radio relic origin for one of the filaments in the north. The rest of the emission could be the result of  turbulence associated with the
primary and a possible secondary merger. 

If a component of the merger along the line of sight was missed, the total intensity emission as well as the new polarized large-scale patches presented here could be compatible with a radio relic generated by a shock wave propagating almost perpendicularly to the plane of the sky towards the observer, or with a small inclination with respect to it. Deeper and more detailed X-ray and optical data are necessary to shed light on this and understand if a radio relic interpretation is valid.

\subsubsection*{Cluster radio halo}

The diffuse source in A523 has been historically classified as a radio halo \citep{Giovannini2011}. A detailed characterization of the properties of the system raised doubts about this classification in the last decade, among them the detection of polarized emission and the offset between the radio and X-rays peak \citep{Girardi2016}. 

Typically, radio halos are not observed in polarization due to the mixing of thermal and emitting radio plasma (internal depolarization) and external beam depolarization and sensitivity issues \citep{Govoni2013}. A523 is one of the few radio halos where total intensity filaments of emission show a polarized counterpart \citep{Girardi2016}.
The A523 galaxy cluster is characterized by a lower X-ray luminosity than expected from the radio power - X-ray luminosity correlation observed for other systems hosting radio halos, indicating a thermal gas density and/or temperature possibly lower than in other galaxy clusters hosting similar sources. A comparison with the X-ray emission reveals that P1, P2, and P3 are located in projection in the outskirts of the cluster, where the thermal gas density is lower and the magnetic field is weaker than in the cluster centre. For these reasons there could be low internal depolarization, favoring the detection of the polarized emission in the frequency range 1-2\,GHz. Similarly, the polarized emission associated with P0 could be due to a peripheral filament of magnetic field seen in projection. This interpretation is supported by the rotation measure analysis. Indeed, the polarization angle versus the squared observing wavelength shows a linear trend, indicating a rotation completely due to a foreground medium, possibly our own Galaxy, as suggested by 
the comparison of the rotation measure properties of the source with the Galactic rotation measure within a radius of 5$^{\circ}$ in the same direction of the sky (see \S\,\ref{Rotation Measure analysis}). 
\begin{table*}
\caption{Adopted parameters in the numerical simulations}  
\label{tab2}      
\centering          
\begin{tabular}{c l l}    
\hline       
    Parameter       & Value                    & Description          \\ 
\hline  
$n_0$               &$1.08\times 10^{-3}$\,cm$^{-3}$                                               & Central gas density\\
$r_{\rm c}$         &188-kpc                                                                    & Core radius\\
$\beta$             & 0.42                                                                          & Slope of the $\beta$-model \\
\hline
z                   & 0.104                                                                         & Redshift of the system\\
\hline
$B_0$               & 0.5\,$\mu$G                                                                   & Central magnetic field strength        \\
$\eta$              & 0.5                                                                           & Slope of the magnetic field radial decrease\\
$\Lambda_{\rm min}$ & 8-kpc                                                                       & Minimum scale of fluctuation of the magnetic field\\
$\Lambda_{\rm break}$ & 1024-Mpc                                                                        & Break scale of fluctuation of the magnetic field\\
$\Lambda_{\rm max}$ & 2048-Mpc                                                                        & Maximum scale of fluctuation of the magnetic field\\
$n_{\rm low}$       & 0                                                                          & Spectral index of the magnetic field power spectrum for $k_{\rm min}<k<k_{\rm break}$\\
$n_{\rm high}$                 & 11/3                                                             & Spectral index of the magnetic field power spectrum for  $k_{\rm break}<k<k_{\rm max}$, \\
                    &                                                                               &here assumed to be Kolmogorov\\
\hline
$\gamma_{\rm min}$      & 3500                                                                          & Minimum relativistic electron Lorentz factor \\
$\gamma_{\rm max}$      & 1.5$\times$10$^{4}$                                                           & Maximum relativistic electron Lorentz factor \\
$\delta$            & 3.4                                                                             & Power-law index of the energy spectrum of \\
                     &                                                                              & the relativistic electrons \\
$K_0$               & Adjusted to guarantee equipartition                                           & Electron spectrum normalization \\ 
                    & at each point of  the computational grid                                      &  \\

\hline                                 
\end{tabular}
\label{sim}
\end{table*}  

Other systems hosting polarized radio halos as A2255 \citep{Govoni2005,Pizzo2011} and MACS J0717.5+3745 \citep{Bonafede2009} do not show polarized emission on such spatial scales, suggesting that in A523 there must be present a magnetic field %fluctuating 
over larger scales than typically observed in galaxy clusters.

\cite{Girardi2016} investigated the intracluster magnetic field properties 
through the comparison of the total intensity and polarized signal of the diffuse emission with synthetic images. They concluded that the radio brightness of the central diffuse source in total intensity and polarization, and its disturbed morphology, are well reproduced by a Gaussian random magnetic field power-spectrum radially decreasing in strength as the square root of the thermal gas density, with a central strength $B_0=0.5\,\mu$G and a maximum fluctuation scale of $\approx$\,1\,Mpc. We note that the value of the
magnetic field strength derived by \cite{Girardi2016} is comparable  
with the lower limit by \cite{Cova2019}, i.e. 0.2\,$\mu$G over the whole radio halo region and 
0.8\,$\mu$G when only the brightest region of the cluster is considered. 

To understand whether this large-scale magnetic field could be responsible for  radio halo emission with the new features presented in this work, we utilize numerical simulations produced by the software \textsc{FARADAY} \citep{Murgia2004}.
According to the dynamo scenario, the magnetic field setup used by \cite{Girardi2016} is not completely realistic since no power is assumed at scales larger than the maximum scale of fluctuation of the magnetic field power spectrum. The consequence is an unnatural morphology of the resulting simulated magnetic field that then exhibits strong anti-correlations, potentially leading to a patchy radio emission.

Even if this effect is partially reduced by the convolution of the magnetic field power spectrum with the thermal gas density (see fig.\,10 \citealt{Vacca2012}), to completely avoid it, we adopt here
a setup similar to that used by \cite{Girardi2016}, 
 but with a double power law power spectrum with slope $n_{\rm low}=0$ for 1\,$\leq \, k_{\rm break}\,<\, 2$ and $n_{\rm high}=11/3$ for 2\,$\leq \,k_{\rm break}\,\leq $\,256. The wave-number $k$ is linked to the spatial scale of fluctuation $\Lambda$ by $\Lambda=N_{\rm pix}\times\Delta/k$, where $N_{\rm pix}$ is the size of the computational grid in pixels and $\Delta$ is the size of a pixel in kpc, assumed here to be respectively 512-pixels and 4-kpc. 
Moreover, we assume $\delta=3.4$ in agreement with the spectral index $\alpha\,=\,1.2$ ($\delta=2\alpha+1$) derived by \cite{Vacca2022}.  
A summary of the parameters adopted for the numerical simulations is given in Table\,\ref{sim}. We converted them according to our cosmology when necessary. Our aim is to look for the presence of significant polarized emission with total intensity counterpart below the 3$\sigma_{\rm I}$. 
In Fig.\,\ref{fig6}, we present the synthetic images in total intensity (top) and polarization (bottom) after convolving with a circular FWHM of 56-arcsec for different realizations of the simulated cubes. The contours in the images are drawn at the 3$\sigma_{\rm I}$ in the top panels and at the 3$\sigma_{\rm P}$ in the bottom panels. These simulations indicate the presence of significant polarization patches above the 3$\sigma_{\rm P}$ with a total intensity counterpart below the 3$\sigma_{\rm I}$. By assuming an upper limit in total intensity of 3$\sigma_{\rm I}=$0.45\,mJy\,beam$^{-1}$ as derived from the observations at 1.410-GHz, we obtain a lower limit in polarization of about 30--33 percent. 

Overall, with a magnetic field fluctuating on scales as large as $\approx$\,Mpc we are able to explain the radio brightness levels and morphology of the central diffuse source in total intensity and polarization as well as to reproduce the presence of polarization patches on larger spatial scales than the total intensity emission, as observed in A523. According to our results, this detection is possible due to a noise level in polarization three times better than in total intensity.

\section{Conclusions}
\label{conclusions}
In this paper we present new polarimetric observations  of the diffuse emission in A523 in the frequency range 1-2\,GHz. Beside the polarized emission associated with the central radio halo,
we find for the first time evidence of polarized emission on scales of 2.5-Mpc. Our analysis suggests that this emission is real: it has been detected with two independent observations and turns out to be unrelated to instrumental polarization. 
By analysing the rotation measure properties of the system and resorting to multi-frequency observations and numerical simulations, we infer that this polarized emission is likely associated with filaments of the radio halo located in the outskirts of the system where the magnetic field is weak and the thermal gas density low, and hence depolarization is expected to be low.

A magnetic field with a central strength of $B_0=0.5\,\mu$G and fluctuating on scales as large as $\approx$\,1\,Mpc is able to explain the radio brightness levels and morphology of the central diffuse source in total intensity and polarization as well as to reproduce the presence of polarization patches on larger spatial scales than the total intensity emission, as observed in A523. These patches can be detected thanks to a noise level in polarization three times better than in total intensity.

\section*{Acknowledgements}
We thank the anonymous referee whose comments helped to improve the quality of the paper.
VV and MM acknowledge support from INAF mainstream project “Galaxy Clusters Science with LOFAR” 1.05.01.86.05. 
FG, MM, and FL acknowledge financial support from the Italian Minister for Research and Education (MIUR), project FARE, project code R16PR59747, project name "FORNAX-B". 
The Enhancement of the Sardinia Radio Telescope (SRT) for the study of the Universe at high radio frequencies is financially supported by
the National Operative Program (Programma Operativo Nazionale - PON) of the Italian Ministry of University and Research "Research and
Innovation 2014-2020", Notice D.D. 424 of 28/02/2018 for the granting of funding aimed at strengthening research infrastructures, in
implementation of the Action II.1 – Project Proposals PIR01\_00010 and CIR01\_00010. The National Radio Astronomy Observatory (NRAO) is a facility
of the National Science Foundation, operated under cooperative agreement by
Associated Universities, Inc.

%%%%%%%%%%%%%%%%%%%%%%%%%%%%%%%%%%%%%%%%%%%%%%%%%%
\section*{Data Availability}

The data underlying this article will be shared on reasonable request to the corresponding author.



\bibliographystyle{mnras}
\bibliography{mnras} % if your bibtex file is called example.bib

%\begin{thebibliography}{99}
%\end{thebibliography}

%%%%%%%%%%%%%%%%%%%%%%%%%%%%%%%%%%%%%%%%%%%%%%%%%%

%%%%%%%%%%%%%%%%% APPENDICES %%%%%%%%%%%%%%%%%%%%%

\appendix

\section{Archival radio data}
\label{appendix_a}

\begin{figure}
	\includegraphics[width=0.5\textwidth]{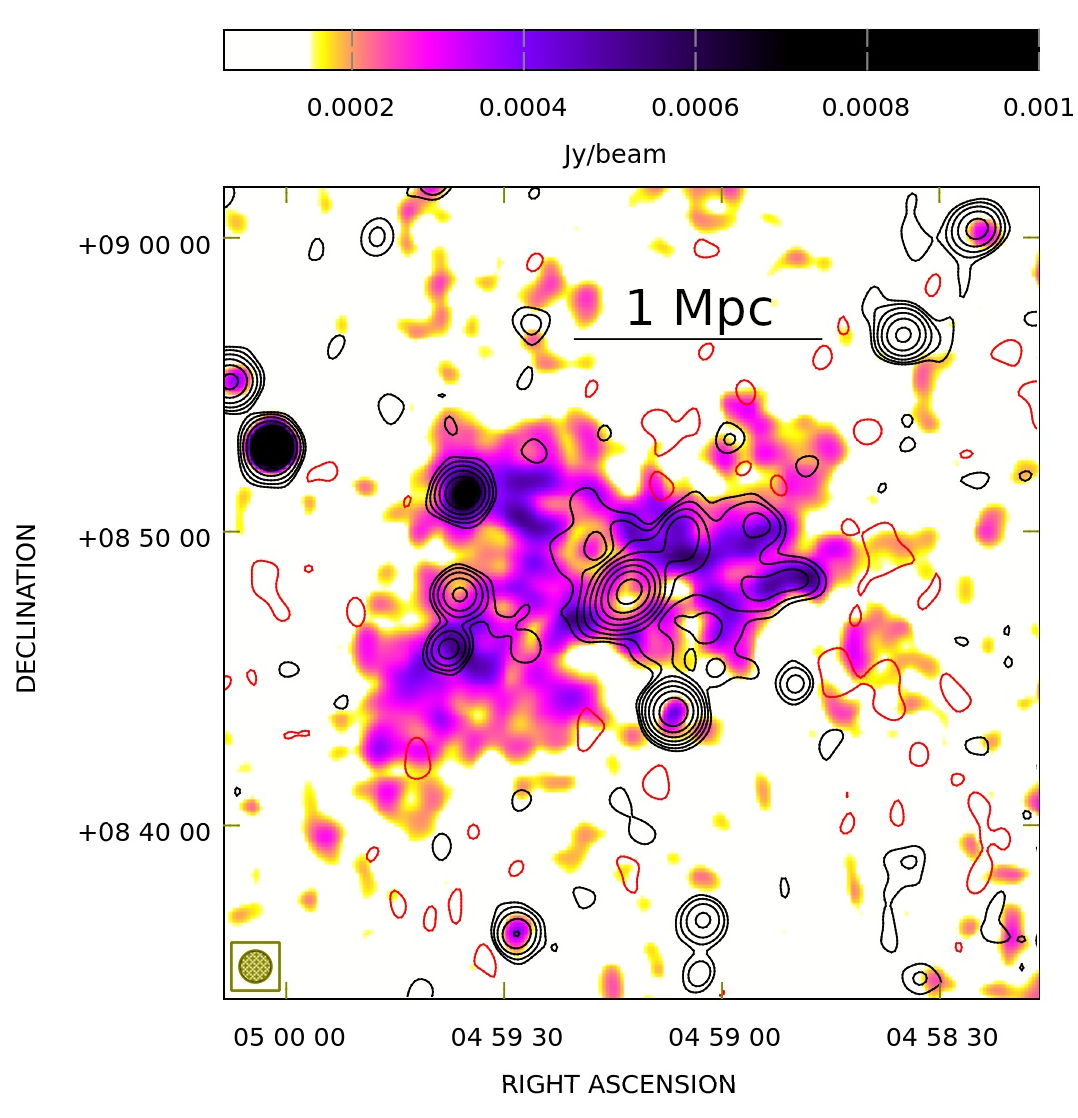}
    \caption{Emission in total intensity (contours) and polarization (colors) of the galaxy cluster A523 from archival VLA data at 1.4-GHz. Black contours start at 3$\sigma_{\rm I}$ ($\sigma_{\rm I}=0.15\,$mJy\,beam$^{-1}$ at 60-arcsec) and scale by a factor 2. Red contours have been drawn at -3$\sigma_{\rm I}$.}
    \label{figA}
\end{figure}

In this Appendix we present archival VLA data (project ID AR690) observed on 2009 December 28 at 1.4-GHz 
in D configuration, with a total bandwidth of 50-MHz in continuum mode and in full Stokes for a total observing time of about 1.43\,h. These data have been presented in \cite{Girardi2016}, after applying a cut at 3$\sigma$ in fractional polarization.  Refer to that paper (section 5.1) for details about the data reduction. In Fig.\,\ref{figA} we present the total intensity and polarized emission obtained with these data using \textsc{CASA} with the
same settings  
used to produce the images in Fig.\,\ref{fig1}. Without applying any cut in fractional polarization, a clear polarized signal is detected in the direction of the cluster and extends beyond the total intensity emission. 
The dataset covers an elevation range from 20$^{\circ}$ to 50$^{\circ}$. The presence of the polarized emission is confirmed also when imaging the elevation ranges 20$^{\circ}$-35$^{\circ}$ and 35$^{\circ}$-50$^{\circ}$ separately (not shown here). 
This excludes shadowing of the VLA antennas as a possible cause of image artifacts. In any event, shadowing of the antennas is not expected to affect the new data presented in \S\,\ref{results}, since they span an elevation range between 60$^{\circ}$ and 66$^{\circ}$.

\begin{figure}
	\includegraphics[width=0.5\textwidth]{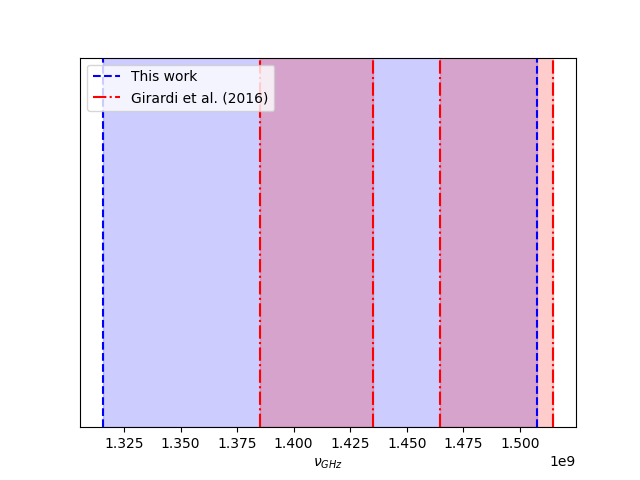}
    \caption{Frequency coverage of the dataset presented in this work in blue and of the dataset presented in \protect\cite{Girardi2016} in red.}
    \label{figA2}
\end{figure}

Comparison of the image in Fig.\,\ref{fig1} with the image in Fig.\,\ref{figA} shows they are
characterized by a different
brightness distribution and size of the polarized emission. 
The two datasets are characterized by a different frequency coverage, as shown in Fig.\,\ref{figA2}. The dataset presented in this work continuously covers the frequency range from 1.3155-GHz to 1.5075-GHz, while the dataset by \cite{Girardi2016} utilizes two windows in the upper part of this frequency range. The combination of an higher central frequency and of a smaller frequency band of the second one could be responsible of a reduced depolarization of the signal, as well as its longer exposure time providing a better uv-coverage, that therefore overall could explain why the source appears brighter and more extended. This is indeed consistent with results from RM synthesis, see Fig.\,\ref{fig4a}.

\section{Point sources embedded in the diffuse emission}
\label{appendix_c}

\begin{figure}
	\includegraphics[width=0.5\textwidth]{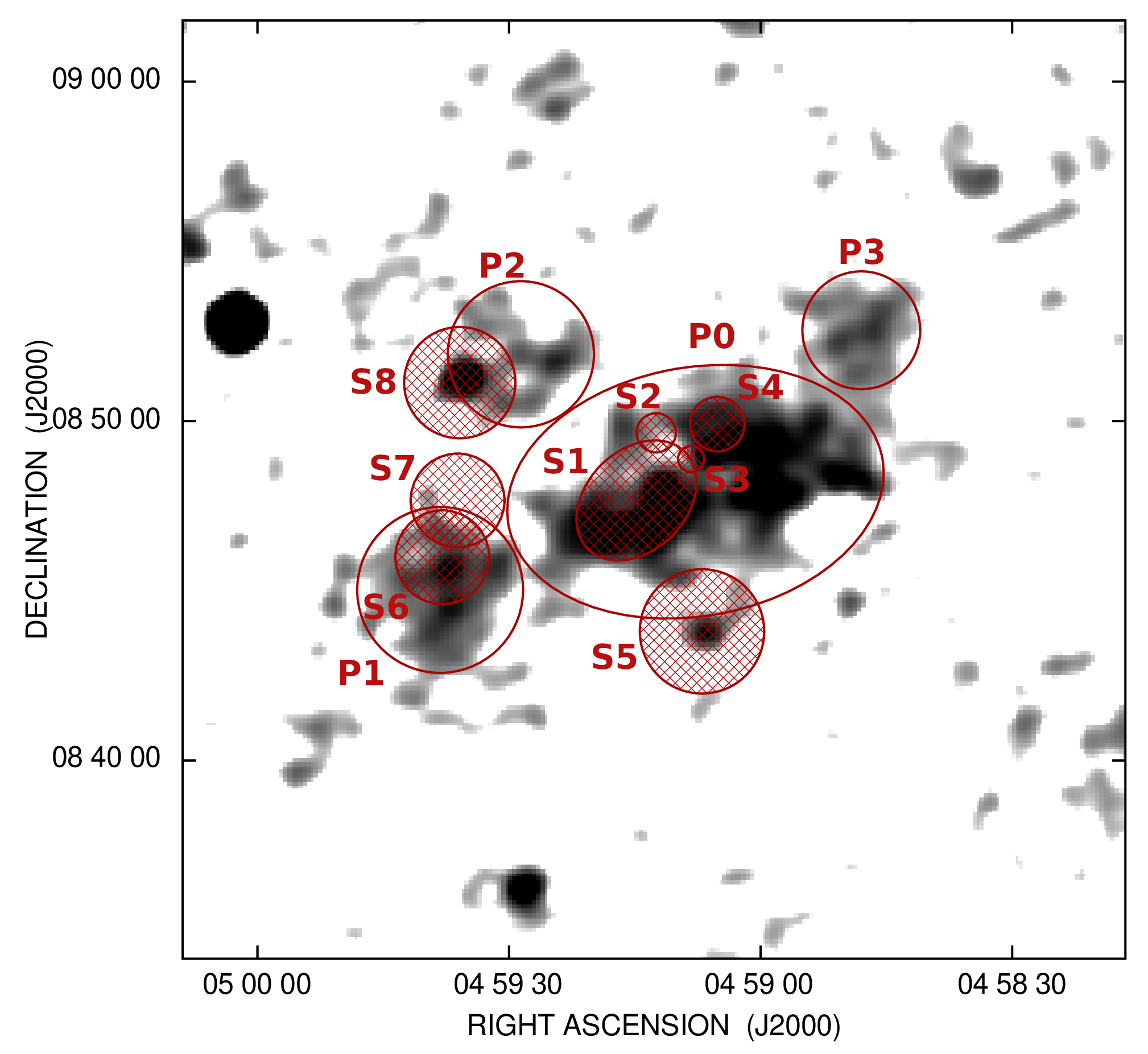}
    \caption{Emission in polarization (grey colors) of the galaxy cluster A523 at 1.410-GHz. Polarization properties of the patches P0, P1, P2 and P3 have been measured using red empty boxes shown in figure, after masking embedded discrete sources S1, S2, S3, S4, S5, S6, S7 and S8, as shown by red filled boxes.}
    \label{figC}
\end{figure}

In Fig.\,\ref{figC} we show the polarized emission above 3$\sigma_{\rm P}$ (corrected for the
positive bias) in the central region of the galaxy cluster A523 at 1.410-GHz in grey colors. The regions used to derive polarization properties of the patches P0, P1, P2 and P3 are shown by red empty boxes. They have been drawn in such a way to include for each patch the emission above 3$\sigma_{\rm P}$ at 1.410-GHz. Possible embedded discrete sources have been masked, as shown by the red filled box in Fig.\,\ref{figC}. These sources have been labelled S1, S2, S3, S4, S5, S6, S7 and S8, following the nomenclature used in \cite{Vacca2022}.

\section{Galactic emission}
\label{appendix_b}

\begin{figure*}
	\includegraphics[width=0.9\textwidth]{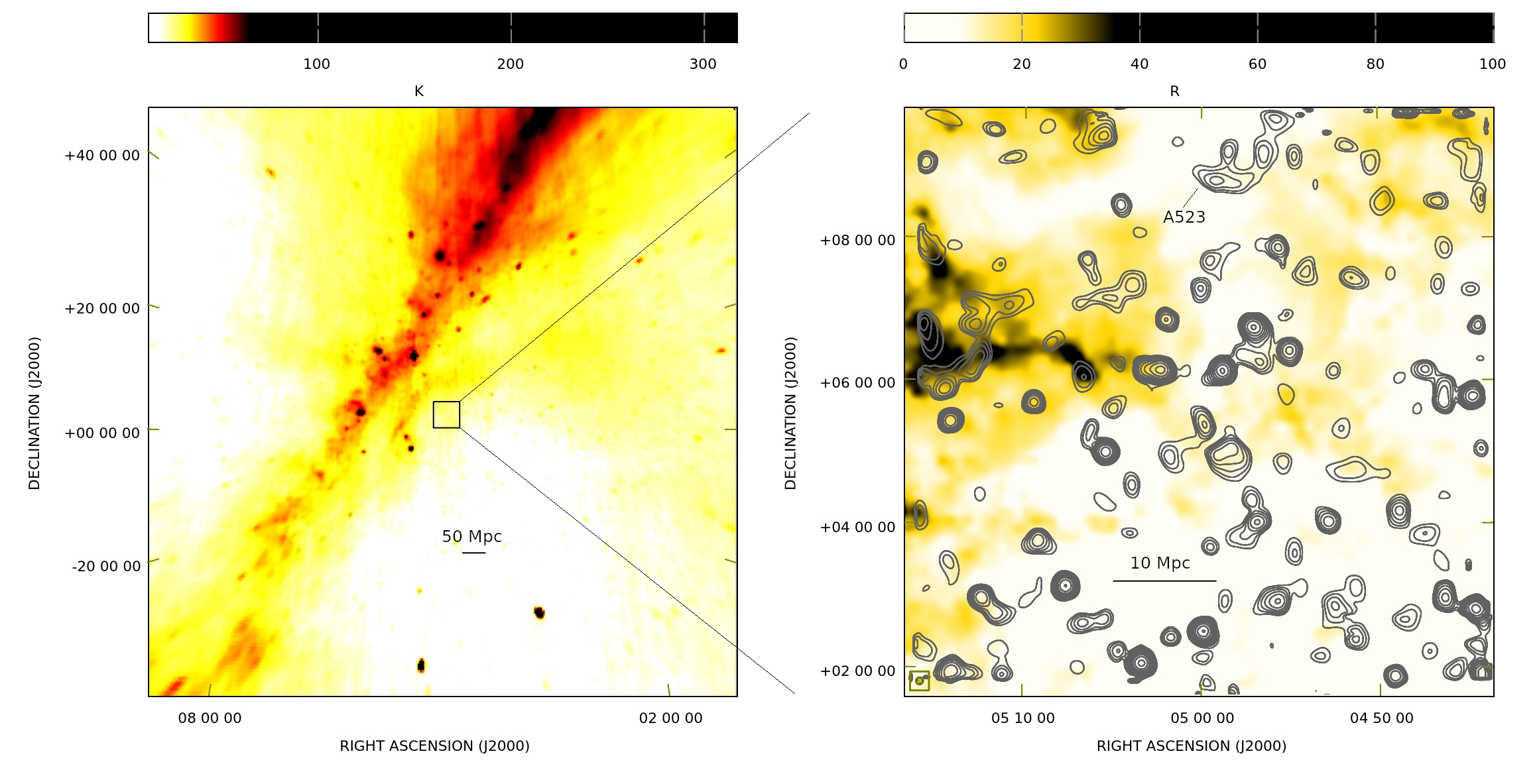}
    \caption{Left: synchrotron emission of the sky in Kelvin at 408-MHz by \protect\cite{Haslam1982}. A523 is located at the centre of the image and its position is marked by a square. The resolution of the image is 0.85$^{\circ}$. % as shown in the bottom left. 
    Right: H-alpha image in Rayleighs \protect\citep[1R=10$^6$/4$\pi$ photons\,cm$^{-2}$\,s$^{-1}$\,sr$^{-1}$, at 6-arcmin FWHM resolution as shown in the bottom left,][]{Finkbeiner2003} overlaid with radio contours at 1.55-GHz from the SRT 
    \protect\citep[see fig.\,2][]{Vacca2018} at a spatial resolution of 13.9\,arcmin$\,\times\,$12.4\,arcmin.}
    \label{figB}
\end{figure*}

In order to investigate a possible Galactic origin of the polarized signal observed in the direction of A523, we examined the presence of large-scale Galactic emission in this region. In Fig.\,\ref{figB} we show on the left the continuum synchrotron emission of the sky at 408-MHz by \cite{Haslam1982}. The poor spatial resolution of this image does not allow us to exclude the presence of Galactic emission at this location. Therefore, on the right panel of Fig.\,\ref{figB} we superimpose the H-alpha image of our Galaxy \citep{Finkbeiner2003} obtained from a combination of the Virginia Tech Spectral in the north and the Southern H-Alpha Sky Survey Atlas in the south with the SRT radio contours of an 8$^{\circ}\times$8$^{\circ}$ region hosting the cluster A523 by \cite{Vacca2018}. The SRT radio contours provide a better view of 
synchrotron sources in the field, while the H-alpha signal on free-free emission. No clear Galactic counterparts can be identified.

%%%%%%%%%%%%%%%%%%%%%%%%%%%%%%%%%%%%%%%%%%%%%%%%%%

% Don't change these lines
\bsp	% typesetting comment
\label{lastpage}
\end{document}